\documentclass[twocolumn]{aa}
\usepackage{epsfig}
\usepackage[latin1]{inputenc}
\def\simlt{\ \raise -2.truept\hbox{\rlap{\hbox{$\sim$}}\raise5.truept   %
\hbox{$<$}\ }}
\def\simgt{\ \raise -2.truept\hbox{\rlap{\hbox{$\sim$}}\raise5.truept   %
\hbox{$>$}\ }}                                                          %

\def\be{\begin{equation}}
\def\ee{\end{equation}}
\def\newline{\hfil\break}

\def\la{\mathrel{\hbox{\rlap{\hbox{\lower4pt\hbox{$\sim$}}}\hbox{$<$}}}}
\def\ga{\mathrel{\hbox{\rlap{\hbox{\lower4pt\hbox{$\sim$}}}\hbox{$>$}}}}

\begin{document}

\title{Disentangling the gamma-ray emission of NGC1275 and that of the Perseus cluster}

   \author{S. Colafrancesco\inst{1,2} P. Marchegiani\inst{2,3} and P. Giommi\inst{1}}

   \offprints{S. Colafrancesco}

   \institute{   ASI-ASDC
              c/o ESRIN, Via G. Galilei snc, I-00040 Frascati, Italy
              Email: colafrancesco@asdc.asi.it
   \and
              INAF - Osservatorio Astronomico di Roma
              via Frascati 33, I-00040 Monteporzio, Italy.
              Email: sergio.colafrancesco@oa-roma.inaf.it
   \and
              Dipartimento di Fisica, Universit\`a di Roma La Sapienza, P.le A. Moro 2, Roma, Italy
              Email: paolo.marchegiani@oa-roma.inaf.it
             }

\date{Received  / Accepted  }

\authorrunning {S. Colafrancesco et al.}

\titlerunning {Gamma rays in the Perseus cluster}

\abstract{The Gamma-ray emission from galaxy clusters hosting
active galaxies is a complex combination of diffuse and point-like
emission features with different spectral and spatial properties.}
{We discuss in details the case of the Perseus cluster containing
the radio-galaxy NGC 1275 that has been recently detected as a
bright gamma-ray source by the {\it Fermi}-LAT experiment, in
order to disentangle the sources of emission.}
{We provide a detailed study of the gamma-ray emission coming from
the core of Perseus by modeling the central AGN emission with a
multiple plasma blob model, and the emission from the extended
cluster atmosphere with both a Warming Ray (WR) model and Dark
Matter (DM) neutralino annihilation models. We set constraints on
both the central galaxy and cluster SED models by using both
archival multi-frequency data and the recent very high energy
observations obtained by Fermi and MAGIC.}
{We find that: i) in all the viable models for the cluster
gamma-ray emission, the emission detected recently by Fermi from
the center of the Perseus cluster is dominated by the active
galaxy NGC 1275, that is found in a high-emission state; ii) the
diffuse gamma-ray emission of the cluster, in the WR model and in
the DM models with the highest allowed normalization, could be
detected by Fermi if the central emission from NGC1275 is in a
low-emission state; iii) Fermi can have the possibility to resolve
and detect the diffuse gamma-ray flux (predicted by the WR model)
coming from the outer corona of the Perseus cluster atmosphere at
distances $r \simgt 800$ kpc.
These results are consistent with the evidence that in the other
frequency bands, the diffuse cluster emission dominates on the
central galaxy one at low radio frequencies with $\nu\simlt1$ GHz
and at X-ray energies of order of $E\sim$ keV.}
{Our results show that a simultaneous study of the various
emission mechanisms that produce diffuse gamma-rays from galaxy
clusters and the study of the emission mechanisms that produce
gamma-rays from active galaxies residing in the cluster
atmospheres is absolutely crucial first to disentangle the
spectral and spatial characteristics of the gamma-ray emission and
secondly to assess the optimal observational strategy in the
attempt to reveal the still elusive diffuse gamma-ray emission
widely predicted for the atmospheres of large-scale structures. }


 \maketitle

\section{Introduction}
 \label{sec.intro}

The radio-galaxy NGC 1275 (3C 84) has been recently detected by
{\it Fermi} as a source of high-energy gamma rays  with an average
flux and power-law photon index of $F(>100 \mbox{ MeV}) = (2.10
\pm 0.23) \times 10^{-7}$ cm$^{-2}$ s$^{-1}$ and $\gamma = 2.17
\pm 0.05$, respectively (Abdo et al. 2009). The emission detected
by the {\it Fermi}-LAT is consistent with a point source centered
at the nucleus of NGC 1275. The gamma-ray flux measured with {\it
Fermi}-LAT is almost an order-of-magnitude brighter than the EGRET
flux upper limit, $F(>100 \mbox{ MeV}) < 3.72 \times 10^{-8}$
cm$^{-2}$ s$^{-1}$ (Reimer et al. 2003), and therefore implies that NGC 1275 is varying
significantly at gamma-rays on time scales from months to years
(see discussion in Abdo et al. 2009).

The {\it Fermi} results on NGC1275 have clearly an impact on the
models for the possible gamma-ray emission originating from the
surrounding Perseus cluster atmosphere because they limit the
detectability of the amount and of the spectral energy
distribution (SED) of the diffuse gamma-ray emission that can
originate from the hosting galaxy cluster.

The Perseus cluster (as many other galaxy clusters) is indeed
expected to be a source of gamma-ray emission due to various
emission mechanisms (see e.g., Colafrancesco 2007-2009 for a
recent discussion):
%
i) Inverse Compton Scattering (ICS) of CMB photons and
relativistic bremsstrahlung of primary cosmic ray particles
(mainly electrons) (see, e.g., Houston et al. 1984, Sarazin 1999,
Miniati et al. 2001, Brunetti 2003, Colafrancesco et al. 2005);
ii) neutral pion e.m. decay produced by pp collisions, and ICS of
CMB photons and relativistic bremsstrahlung of secondary electrons
produced in the same pp collisions (e.g., Dennison 1980,
Colafrancesco \& Blasi 1998, Marchegiani et al. 2007);
iii) neutral pion e.m. decay produced by Warming Rays (WR)
interaction and ICS of CMB photons and relativistic bremsstrahlung
of secondary electrons produced in the same pp collisions
(Colafrancesco and Marchegiani 2008, 2009);
iv) neutral pion e.m. decay produced by neutralino Dark Matter
annihilation, and ICS of CMB photons and relativistic
bremsstrahlung of secondary electrons produced in the same DM
annihilation processes (Colafrancesco \& Mele 2001, Colafrancesco
et al. 2006, 2010);
v) ICS and relativistic bremsstrahlung from PeV electron-positron
pairs produced via interactions of Ultra High Energy photons
(particles) with the CMB photons (Timokhin et al. 2004, Inoue et
al. 2005).

Experimental limits on the gamma-ray emission from the Perseus
cluster and its central galaxy NGC1275 have been obtained by
various experiments.
Gamma-ray observations toward NGC 1275 and the Perseus clusters
were first reported in the 1980's by Strong \& Bignami (1983). The
COS B data, taken between 1975-1979 (Strong et al. 1982;
Mayer-Hasselwander et al. 1982), show a gamma-ray excess at the
position of the galaxy, although evidence for emission uniquely
related to NGC 1275 is ambiguous (in fact positional uncertainties
were not given for the COS B data). Interpreted as emission from
NGC 1275, the gamma-ray flux is $F (> 70 \mbox{ MeV}) = 8.3 \times
10^{-7}$ cm$^{-2}$ s$^{-1}$, which is more than an order of
magnitude higher than the upper limit reported by EGRET.
Beyond the {\it Fermi} detection and the EGRET upper limit already
mentioned, recent limits on TeV emission at $E>400 $ GeV have been
obtained by Whipple ($F<7.4 \times 10^{-12}$ erg cm$^{-2}$
s$^{-1}$; see Perkins et al 2006) and from MAGIC at $E> 100$ GeV
($F< 4.6$ to $7.5$ $\times 10^{-12}$ cm$^{-2}$ s$^{-1}$, for
spectral indices in the range $-1.5$ to $-2.5$; see Aleksic et al.
2009).
Very recently, a new upper limit has been obtained with VERITAS
(Acciari et al. 2009), with $F(>188 \mbox{ MeV}) <5.11 \times
10^{-12}$ cm$^{-2}$ s$^{-1}$ while NGC1275 was in a quite low
state at the gamma-ray energies probed by {\it Fermi}.\\
At soft gamma-ray energies, INTEGRAL observations of the Perseus
cluster found no evidence for a diffuse power-law emission which
would dominate the emission above 30 keV (Eckert \& Paltani 2009).
However, the angular resolution of IBIS/ISGRI is not sufficient to
disentangle the point-like emission from the diffuse emission
component, so it is not possible to set any definite upper limit
on the diffuse non-thermal emission from the cluster.\\
At X-ray energies, Perseus is the nearest example of a
proto-typical cool core cluster in which the intra-cluster (IC)
gas temperature decreases from the outer region value of $\sim7$
keV down to the value of $\sim 3$ keV found at its center (see
Colafrancesco \& Marchegiani 2008 and references therein for a
recent discussion) where the cluster hosts the giant elliptical
galaxy NGC 1275.\\
The Perseus cluster appears, moreover, to contain a non-thermal
component, namely an excess of hard X-ray emission above the
thermal bremsstrahlung from the diffuse hot cluster gas. Based on
a deep Chandra observation, the non-thermal X-ray component has
been mapped over the core of the cluster and shows a morphology
similar to the radio mini-halo (Sanders, Fabian \& Dunn 2005;
Sanders \& Fabian 2007). Notice, however, that this claim was
questioned on the basis of a long XMM-Newton exposure (Molendi \&
Gastaldello 2009). Above 10 keV, a hard X-ray component has been
detected with HEAO-1 (Primini et al. 1981) and BeppoSAX/PDS
(Nevalainen et al. 2004), although it was not detected with
CGRO/OSSE in the $0.05-10$ MeV range (Osako et al. 1994). It is
interesting to notice that Perseus is the only cluster - out of
ten clusters detected in the $15-55$ keV range with Swift/BAT
(Ajello et al. 2009) - that displays a high-energy non-thermal
component up to 200 keV, even though the hard tail seen in the BAT
spectrum is likely due to nuclear emission from NGC 1275 rather
than to non-thermal emission from the cluster atmosphere. This
idea is supported by possible flux variations compared to past
hard X-ray observations, and by the fact that the extrapolation of
the BAT spectrum is in good agreement with the luminosity of the
nucleus as measured with XMM-Newton (Churazov et al. 2003).\\
This supports, once more, the interpretation that the non-thermal
emission of hard X-rays and of the gamma-rays is produced by the
central active galaxy.

{NGC 1275 has been variously classified as a Seyfert 1.5 (because
of its emission-line optical spectrum, where broad lines are
detected, see V\'eron-Cetty \& V\'eron 1998), but also as a blazar
(due to the strong and rapid variability of the continuum emission
and its polarization, see e.g., Angel \& Stockman 1980; see also
Pronik, Merkulova \& Metik 1999).
The very bright radio source Perseus A (also known as 3C 84) found
in NGC 1275 has a strong, compact nucleus which has been studied
in detail with VLBI (Vermeulen et al 1994, Taylor \& Vermeulen
1996, Walker et al. 2000, Asada et al. 2006) and a bowshock-like
southern jet component (Kellermann et al. 2004; Lister et al.
2009). The radio emission continues on larger scales, and shows a
clear interaction with the intra cluster (IC) gas. Observations
with ROSAT (Boehringer et al. 1993) and later Chandra (Fabian et
al. 2003a, 2006) reveal the presence of cavities in the IC gas,
suggesting that the jets of 3C 84 have blown multiple bubbles in
the atmosphere of the Perseus cluster. On even larger scales,
Perseus exhibits a mini-halo of size $\sim 300$ kpc, best seen at
low-frequency radio emission (Burns 1990), that is likely produced
by synchrotron emission from widely distributed relativistic
particles and fields energized in the central regions of the
cluster.

On the theoretical side, several predictions on the intensity and
spectral properties of the Perseus cluster gamma-ray emission have
been presented.
Pfrommer (2008) predicted a total gamma-ray flux at $E>100$ MeV
that is in the range $(3.2 - 5.6) \times 10^{-9}$ cm$^{-2}$
s$^{-1}$ depending on the details of their models of relativistic
electrons accelerated at cosmological structure formation shocks
and those that are produced in hadronic interactions of cosmic
rays with ambient gas protons.
Kushnir \& Waxman (2009) predicted, in a simple model that
explains the HXR emission from galaxy clusters as ICS scattering
of CMB photons by relativistic electrons accelerated at the
accretion shock surrounding the cluster, a gamma-ray flux at $E>
50$ GeV for Perseus (within 0.1 deg) of $3.1 \times 10^{-13}$
cm$^{-2}$ s$^{-1}$ for the pion decay channel production, and of
$1.5 \times 10^{-13}$ cm$^{-2}$ s$^{-1}$ for the ICS channel
production.
Colafrancesco \& Marchegiani (2008) predicted that the simplest WR
model should produced a total (maximal) gamma-ray flux for Perseus
(integrated within its virial radius) of $1.2 \times 10^{-8}$
cm$^{-2}$ s$^{-1}$ for photons with  $E> 100$ MeV.

The predictions of the various models have now to contend with the
new observation of the central part of the Perseus cluster
atmosphere where the NGC 1275 gamma-ray emission results dominant.
Furthermore, the understanding of the nature of the central radio
galaxy SED (and especially of its high-E branch at $E \simgt$ GeV)
can help in setting more precise constraints on the amount of
diffuse gamma-ray emission of the Perseus cluster, and on the SED
of the cluster non-thermal emission models, and can help also in
outlining the optimal strategy for the detectability of galaxy
clusters at gamma-ray energies.

In this paper we study the gamma-ray emission of the composite
system of the radio galaxy -- cluster that have different origins
and hence different spectral, temporal and spatial
characteristics. We perform this study with the aim to disentangle
the diffuse and point-like emissions by using a full
multi-frequency strategy. To this aim, we discuss in Sect.2 the
theory of gamma-ray emission in systems composed by a galaxy
cluster plus a central AGN. In Sect. 3 we discuss a model for the
multi-frequency SED of NGC 1275 that is able to recover the
different sets of observations for this radio galaxy, from radio
to high-energy gamma-rays and TeVs. Sect.4 presents the
predictions for the diffuse emission of the atmosphere of the
Perseus cluster in the region surrounding NGC 1275 for three
different physical models: i) the Warming ray model, whose
multi-frequency SED is based on its ability to reproduce the
temperature profile of the hot intracluster gas; ii) three models
for neutralino Dark Matter annihilation with different neutralino
mass and composition; iii) a pure leptonic model with a power-law
relativistic electron spectrum. In Sect.5 we will discuss the
comparison of the SEDs for NGC 1275 and for the surrounding
Perseus cluster atmosphere using the largest possible frequency
coverage, from radio to TeV energies. We will show that a full
multi-frequency, and multi-component study, is able to disentangle
the point-like vs. diffuse emissions of such complex astrophysical
systems. The final Sect.6 is devoted to the discussion of our
results and to summarize our conclusions.

Throughout the paper, we use a flat, vacuum--dominated
cosmological model with $\Omega_m = 0.3$, $\Omega_{\Lambda} = 0.7$
and $h = 0.7$.

\section{The SED of a cluster containing a central radio galaxy: setting the model}

The evidence so far discussed indicates that in order to fully
understand the level and the spectral properties of the high-E
emission from the center of the Perseus cluster, we need to model
in details the two separate components, i.e. NGC1275 and the
Perseus cluster atmosphere.

The model we discuss here consists of a diffuse plasma component,
associated to the atmosphere of the Perseus cluster, and a compact
central component, mainly provided from the jets and blobs emitted
from the NGC 1275 galaxy. We want to compare the two different
SEDs in the same spatial region, i.e. in the central part of the
cluster.\\
In this region we expect most of the diffuse emission produced by
the cluster, since the thermal bremsstrahlung emissivity is
$\propto n_{th}(r)^2$, the non-thermal cluster emission produced
by secondary electrons is $\propto n_{th}\times n_{CR}$, and the
DM-produced emission is $\propto n_{DM}^2(r)$, usually peaked
towards the cluster center (here $n_{th}, n_{CR}$ and $n_{DM}$ are
the number densities of thermal particles, cosmic rays and DM
particles, respectively).
To model the Perseus SED we consider, specifically, the WR model
(Colafrancesco \& Marchegiani 2008), in which a population of
relativistic protons heat the gas due to Coulomb and hadronic
interactions bringing it to a quasi stationary cooling-heating
equilibrium and, at the same time, produce via p-p collisions
secondary electrons and direct gamma-ray emission (via neutral
pion decay). Contrary to standard secondary electron models (see,
e.g., Marchegiani et al. 2007) in which the CR density (and hence
the intensity of the produced radiation) is not known a priori,
the CR density and its spatial distribution in the WR model are
fixed by the condition of reproducing the cluster gas temperature
distribution.\\
Dark Matter annihilation in the central region of the cluster is
also the dominant source of emission in neutralino DM models:
therefore, we consider the SEDs predicted in a few neutralino DM
models whose intensity is, however, uncertain due mainly to the
unknown value of the annihilation cross-section (see, e.g., the
discussion made by Colafrancesco \& Marchegiani 2009).\\
We also discuss the emission features of a pure leptonic model, in
which a diffuse population of relativistic electrons
produce $\gamma$-ray emission in the cluster center:
this case will be used to set the maximum density of non-thermal
electrons allowed by the gamma-ray upper limit measured by EGRET.

As for the active galaxy NGC 1275, we describe its SED as produced
in a model (hereafter referred to as the Cannon Model or CM model)
with multiple components (Colafrancesco \& Marchegiani 2010, in
preparation), finding the optimal jet-blob structure that is able
to reproduce the observed multi-frequency SED. Specifically, we
assume in our CM model that different blobs of relativistic plasma
are emitted at different times from the central source (close to
the hypothetical SMBH). This model predicts that the AGN SED is
the combination of the SSC emissions of each separate blob and the
ICS emission that each blob produces on the photons emitted from
external radiation fields, including those emitted from the SSC
part of each emitted blob. A more simplified version of this model
has been already used to describe the SED of the blazar S05
0716+714 (see Giommi et al. 2008, Vittorini et al. 2009).

We describe, in the following, the specific SEDs of the NGC1275
galaxy of the Perseus cluster separately and then the combination
of the two SEDs in the same spatial region.

\section{The SED of NGC 1275}

Fig. \ref{sed_1275} shows the fit to the multi-$\nu$ SED of NGC
1275 obtained in the CM model (see Colafrancesco \& Marchegiani
2010, in preparation) with the composition of three SSC
components, whose parameters are reported in Tab. \ref{tab.1},
generated by three separated plasma blobs blasted away from the
inner regions of the NGC 1275 nucleus.
The first blob of intermediate energy and larger radius produces a
SSC emission that recovers the low-$\nu$ radio emission and the
synchrotron peak and, in its inverse Compton scattering (ICS)
branch, also the historical X-ray data with lower flux.
The second, most energetic and smaller blob produces a SSC
emission that is systematically displaced towards high frequencies
and fits the {\it Fermi} data while remaining subdominant at all
other frequencies.
The third, less energetic blob produces a SSC emission that is
able to fit the X-ray data of higher flux with its IC peak
emission but remains sub-dominant at all other frequencies.
The three components are interpreted as emissions from blobs
blasted away at different times so that their combination could
explain also the source variability.
Tab. \ref{tab.1} reports the values of the boosting parameter
$\delta=[\Gamma(1-\beta \cos \theta)]^{-1}$ (here $\beta=v/c$ is
the bulk velocity of the jet, $\Gamma=(1-\beta^2)^{-1/2}$ and
$\theta$ is the angle between the direction of the jet and the
line of sight) that take values between 4 and 8 for the three
blobs. These values are lower than the typical values of order of
a few tens found for blazars, consistently with the fact that in
the radio-galaxy NGC 1275 the jet is not oriented towards the
observer.\\
If we assume, for the sake of discussion, a viewing angle
$\theta\leq 1/\delta_{max}=0.125$, e.g., $\theta\sim 10^{-1}$ rad,
the values of $\delta$ are compatible with values of the blob bulk
velocity of $\Gamma\sim$ 2--6, for which the emitted radiation is
beamed at angles $\theta\simeq 1/\Gamma=0.17-0.5$ rad. A viewing
angle of 0.1 rad is hence inside the viewing cone within which the
emission is beamed, even though this angle is larger and the bulk
velocity is lower than the values typically found in blazars, in
agreement with the fact that NGC 1275 is not a canonical blazar,
but shows blobs ejected with a viewing angle (w.r.t. the line of
sight) relatively large (but still not of order of $\sim 90$ deg,
as for prototypical radio galaxies) that hence determines emission
characteristics intermediate between a radio galaxy nucleus and
those of a blazar nucleus.
\begin{figure}[ht]
\begin{center}
 \epsfig{file=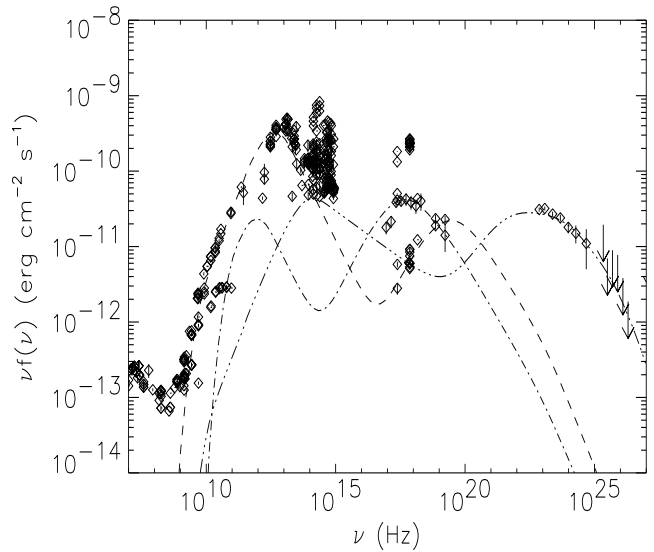,height=8.cm,width=9.cm,angle=0.0}
\end{center}
 \caption{\footnotesize{The SED of NGC 1275 fitted by our
 CM model with three separated SSC components (see Tab.\ref{tab.1}):
 component 1 (dashed), 2 (three dots - dashes) and 3 (dot-dash).
 Data are from NED, from {\it Fermi} (Abdo et al. 2009) and from MAGIC
(Aleksic et al. 2009).
 }}
 \label{sed_1275}
\end{figure}

\begin{table*}[htb]{}
\vspace{2cm}
\begin{center}
\begin{tabular}{|*{9}{c|}}
\hline
 Component & $\log N(\mbox{cm}^{-3})$ & $p_1$ & $p_2$ & $\log \gamma_b$ & $B$ (G) & $\delta$ & $r$ (pc) & $z$\\
\hline
1 & -1.15 & 0.95 & 4.4 & 3.2 & 0.1 & 4.0 & 0.21  & 0.0179\\
2 & 0.83 & 1.2 & 3.5 & 3.7 & 0.1 & 8.0 & 0.008 & 0.0179\\
3 & 3.2  & 1.0 & 4.4 & 2.7 & 0.1 & 8.0 & 0.003 & 0.0179\\
\hline
 \end{tabular}
 \end{center}
 \caption{\footnotesize{Parameters of the three blobs of the CM model whose SEDs are shown
 in Fig.\ref{sed_1275}.
 }}
 \label{tab.1}
 \end{table*}
\begin{figure}[ht]
\begin{center}
 \epsfig{file=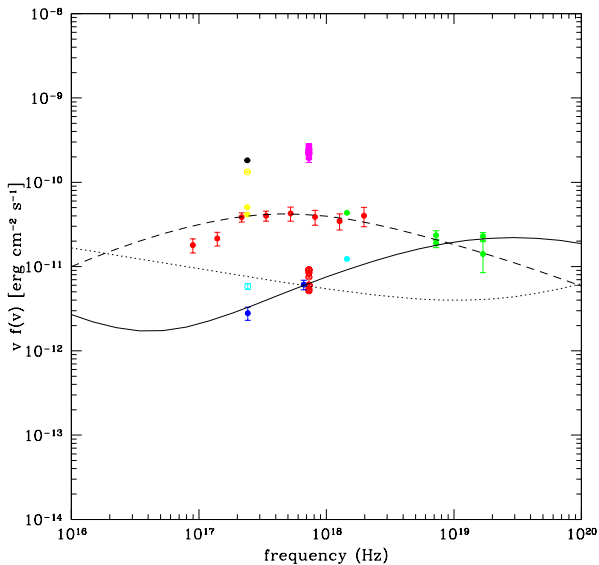,height=8.cm,width=8.cm,angle=0.0}
\end{center}
 \caption{\footnotesize{The multiple-component SED of NGC1275 in the UV to hard
 X-ray frequency range. The three SED components have parameters
 listed in Tab.\ref{tab.1}: component 1 (solid line), 2 (dotted line) and 3 (dashed line).
 Data are from: RASS (black), Einstein IPC (yellow), BeppoSAX WFC (magenta), Swift XRT (red),
 Chandra (blue), XMM (cyan), Integral (green).
 }}
 \label{sed_1275_3c_xray}
\end{figure}

X-ray observations of NGC1275 over a wide historical period show
that the flux of this sources varied substantially (over almost
two orders of magnitude) as well as its spectral shape, indicating
the presence of different SED components.\\
The X-ray part of the SED of NGC1275 (see Fig.
\ref{sed_1275_3c_xray}) shows, in fact, that three blobs SEDs are
required to fit the various sets of historical data accumulated
for NGC1275.\\
In the soft X-ray band, the lowest flux component includes Chandra
data (Balmaverde, Capetti \& Grandi 2006; blue points), the
Swift-XRT data of the lowest stage (data from ASDC; red empty
dots), the XMM data that show a slightly higher flux with respect
to the previous data (Panessa et al. 2006; Evans et al. 2006; cyan
points).\\
The intermediate flux data are from Swift XRT (red filled points),
from the Einstein IPC (data from ASDC; yellow filled points) and
the observations of the Integral JEM-X from Beckmann et al. (2006;
green points).\\
The highest flux component comes from RASS data (data from ASDC;
black points), from the Einstein IPC Slew Survey (data from ASDC;
yellow empty point) and from the BeppoSAX WFC (data from ASDC;
magenta points).
Note that the BeppoSAX WFC data with a spatial resolution of $\sim
5$ arcmin do not resolve the galaxy in X-rays and are likely
contaminated by the underlying cluster emission; the same probably
occurs also with the IPC Slew Survey data and with the RASS data.
Note also that Integral JEM-X has a spatial resolution of $\approx
3.5$ arcmin that does not allow to resolve the central galaxy.\\
In the hard X-ray band, there are data obtained from two different
sets of Integral ISGRI observations: the 2006 observation
(Beckmann et al. 2006; green points) that have a rising spectral
shape which well matches with the low-stage component seen in the
soft X-rays,  and the 2007 observation (Bird et al. 2007; green
points), that have a spectral shape which well matches the
intermediate-stage spectral component seen in the soft X-rays.\\
The SEDs relative to the blobs 1 and 3 of our CM model for NGC
1275 reproduce very well the X-ray data in the low and
intermediate stages, while this model is not able to reproduce the
X-ray data in the highest stage  (i.e. the BeppoSAX WFC, RASS and
Einstein IPC Slew Survey data). These last data could either
indicate the presence of a further component in the SED of NGC
1275 or, more likely (as we will discuss in the next Section), are
contaminated by the unresolved contribution from the thermal
bremsstrahlung emission coming form the intra-cluster gas of the
Perseus cluster.

\section{The diffuse emission of the Perseus cluster}

We describe in this Section the diffuse emission of the Perseus
cluster atmosphere as produced by the combination of thermal
bremsstrahlung from the intracluster gas and non-thermal emission
from  two models, i.e. the Warming Ray (WR) model and a Dark
Matter (DM) annihilation model. We also discuss the case of a pure
leptonic model at the end of this section.

\subsection{The WR model in Perseus cluster}

The WR model (see Colafrancesco \& Marchegiani 2008,
Colafrancesco, Dar \& DeRujula 2004  for details) assumes the
presence of cosmic ray protons in the atmosphere of the Perseus
cluster and that these protons are responsible for all the heating
necessary to quench the central cooling flow.\\
The high energy WR protons can have different origins as has been
discussed in the past: they can be injected by AGN jets
penetrating the cluster atmosphere where they further diffuse and
equilibrate (see Colafrancesco \& Marchegiani 2008 for a
discussion); they can be accelerated by accretion and/or merging
shocks and by intracluster medium (MHD) turbulence (e.g.,
Colafrancesco \& Blasi 1998, Blasi \& Colafrancesco 1999, Brunetti
et al. 2009, Wolfe et al. 2008); they can be accelerated by
supernova remnants (Voelk et al. 1996) or by galactic cannonballs
(Colafrancesco, Dar and DeRujula 2004) occurring in cluster
galaxies and then diffusing and equilibrating in the cluster
atmosphere.\\
The assumption that WR provide all the non-gravitational heating
required to quench cooling flows maximizes the amount and the role
of WR protons in Perseus. However, this provides, clearly, an
extreme case if other heating mechanisms (like, e.g., AGN jets and
lobes, pressure waves, buoyant bubbles and cavities, intra-cluster
shock waves, thermal conduction, leptonic and hadronic
cosmic-rays) are at work in the same place of Perseus. For this
reason, the results we present here for the case of the WR model
can be considered as an upper limit case, even though extremely
useful to set constraints to this model and to the amount of
diffuse gamma-ray emission due to the Perseus cluster.\\
In the WR model that we work out here, we assume a spherical
symmetry for the aims of the present discussion. In such a
theoretical description, we compare our predictions with the
azimuthal averaged density and temperature profiles derived from
X-ray observations. In such a context, the WR model allows to
obtain quantitative results by using an analytical description
that retains all the essential features of the Perseus cluster
structure.\\
It is, nonetheless, clear that in many clusters (and in particular
in Perseus) we observe specific temperature and density structures
that do not follow a spherical symmetry, therefore showing the
complex combination of spatial regions in which heating mechanisms
are more efficient and others in which cooling dominates (e.g.
Fabian et al. 2003b).\\
The fine-grained description of these effects requires, however, a
detailed numerical treatment that goes far beyond the scopes of
our work.

The proton spectrum that we use for WRs in Perseus is
\begin{equation}
N_{WR}(E,r)=N_{WR,0} (E/\mbox{GeV})^{-s} [g_{th}(r)]^{\alpha} \; ,
 \label{wr_spectrum}
\end{equation}
with $E_{min}= m_p c^2 \cdot [1+3.4 \cdot 10^{-5}(kT/\mbox{keV})]$
(see, e.g. Furlanetto \& Loeb 2002) and $E_{max}\rightarrow
\infty$.

The spectral index $s$ is typically in the range 2.3--3.3 (see
discussion in Marchegiani et al. 2007). For the specific case of
Perseus, the available measurements of the mini radio halo
spectrum indicate a radio (synchrotron) spectral index
$\alpha_R\sim1.3-1.4$ (Gitti et al. 2002) that, hence, suggest to
choose a proton spectrum index $s=2.7$.

The function $g_{th}(r)$ describes the radial profile of the IC
gas. Churazov et al. (2003) provided a fitting formula that
describes quite well both the IC gas density and the IC gas
temperature outside the galaxy NGC 1275 and out to the radius $R
\sim 215 \, h_{70}^{-1}$ kpc. Since this radius is of the same
order of magnitude of both the cluster core radius ($\sim200$ kpc;
Churazov et al. 2003) and of the radius of the mini radio halo of
Perseus ($\sim300$ kpc; Burns 1990, Gitti et al. 2002), we assume
here this value} as the radius of the cluster emission region.

In the WR model the density $N_{WR,0}$ and the exponent $\alpha$
that fixes the radial distribution of non-thermal protons are
determined from the request to reproduce the observed distribution
of the cluster temperature by balancing, at each radius, the IC
gas bremsstrahlung cooling and the heating produced by the WRs
mainly through Coulomb and hadronic interactions (see details in
Colafrancesco \& Marchegiani 2008, and  Colafrancesco, Dar \&
DeRujula 2004). The values we obtain for Perseus are:
$N_{WR,0}=1.1\times10^{-7}$ GeV$^{-1}$ cm$^{-3}$ (that corresponds
to a pressure ratio $P_{WR}/P_{th}\sim0.54$ at the cluster center)
and $\alpha=0.91$.

\begin{figure}[ht]
\begin{center}
 \epsfig{file=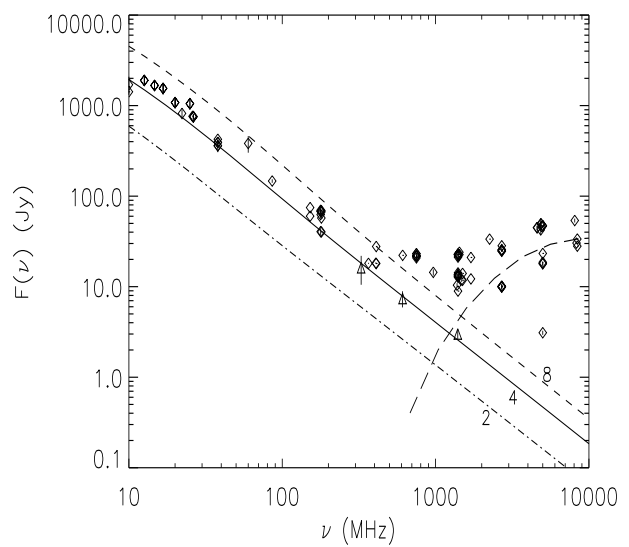,height=8.cm,width=8.cm,angle=0.0}
\end{center}
 \caption{\footnotesize{Radio emission from secondary electrons in the WR model with a
 uniform magnetic field of 2, 4 and 8 $\mu$G.
 Long dashed curve is the SSC model for the galaxy NGC 1275
 (component 1 in Tab. \ref{tab.1}).
 Data are from Gitti et al. (2002) (triangles) and from NED
(diamonds).
 }}
 \label{radio_wr}
\end{figure}

\begin{figure}[ht]
\begin{center}
 \epsfig{file=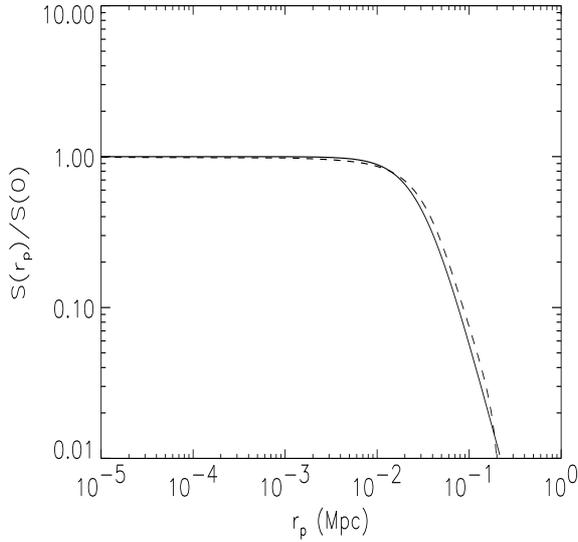,height=8.cm,width=8.cm,angle=0.0}
\end{center}
 \caption{\footnotesize{Radial profile of the synchrotron radio
 emission from secondary electrons in the WR model (dashed curve)
 compared to the best-fit model of the radio-halo brightness profile at
 1.4 GHz derived from Pfrommer \& En\ss lin (2004) (solid curve).
 }}
 \label{radiobril_wr}
\end{figure}

Fig. \ref{radio_wr} shows the radio synchrotron spectrum produced
by secondary electrons in the WR model for a few values of the
magnetic field (here we assume a uniform magnetic field). Fig.
\ref{radiobril_wr} shows the comparison between the best-fitting
curve of the radio brightness at 1.4 GHz found by Pfrommer \&
En\ss lin (2004) with the brightness profile obtained in the WR
model for a uniform magnetic field profile. We can conclude that a
WR model with a uniform magnetic field of $\sim 4$ $\mu$G is in
quite good agreement with the observed radio brightness profile of
Perseus.

\begin{figure}[ht]
\begin{center}
 \epsfig{file=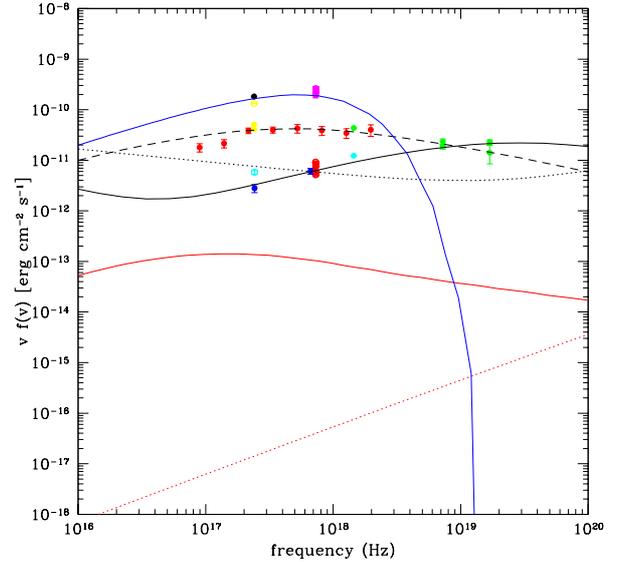,height=8.cm,width=8.cm,angle=0.0}
\end{center}
 \caption{\footnotesize{The HXR ICS-on-CMB emission (red solid curve) and
 bremsstrahlung emission (red dotted curve) of secondary
 electrons in the WR model are compared to the thermal
 bremsstrahlung emission of the IC gas (blue solid curve) and to
 the three separate components of the SSC model for NGC 1275 (see
 Tab. \ref{tab.1}): component 1 (black solid curve), 2 (dotted
 curve) and 3 (dashed curve). Data are taken from NED (see caption of
Fig.\ref{sed_1275_3c_xray} for details).
 }}
 \label{hxr_wr}
\end{figure}

Fig. \ref{hxr_wr} shows the ICS-on-CMB HXR emission and the
non-thermal bremsstrahlung emission of the secondary electrons in
the WR model compared to the thermal bremsstrahlung emission of
the IC gas in the core of Perseus.
The dominant non-thermal emission is given by ICS-on-CMB and it
overcomes even the thermal bremsstrahlung emission at  $E>40$ keV.
The X-ray flux produced by ICS-on-CMB emission in the energy band
2--10 keV is $1.5\times10^{-13}$ erg cm$^{-2}$ s$^{-1}$, while the
flux attributed to the non-thermal component by the analysis of
Sanders et al. (2005) is $6.3\times10^{-11}$ erg cm$^{-2}$
s$^{-1}$; it is clear that such an observed emission cannot be
provided by the WR model.
We must notice here (as already stressed in Sect.2) that the
result of Sanders et al. (2005) has been not confirmed by other
independent observations (Molendi \& Gastaldello 2009, Eckert \&
Paltani 2009, Ajello et al. 2009).

\begin{figure}[ht]
\begin{center}
 \epsfig{file=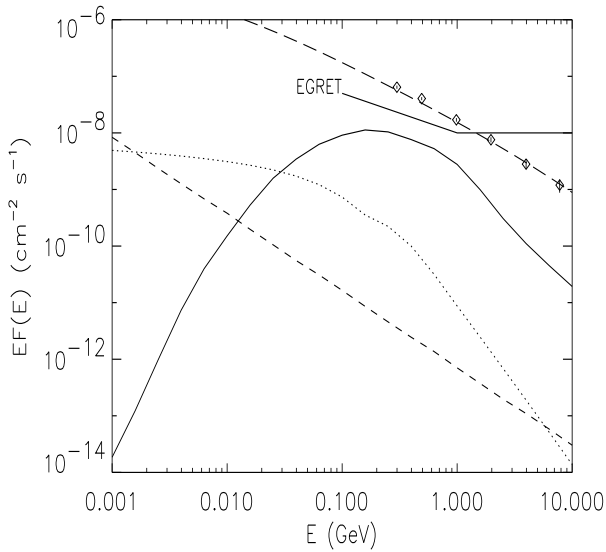,height=8.cm,width=8.cm,angle=0.0}
\end{center}
 \caption{\footnotesize{The gamma-ray emission expected in the WR
 model from the $\pi^0 \to \gamma \gamma$ (solid line) and from
 secondary electrons via ICS-on-CMB emission (short dashes line) and
 non-thermal bremsstrahlung (dotted line).
 The long-dashes line is the SSC model for NGC 1275 (component 2 in Tab. \ref{tab.1}).
 {\it Fermi} data are from Abdo et al. (2009), while the EGRET sensitivity curve
 corresponds a sensitivity limit of 5$\sigma$ in 1 year of observation.
 }}
 \label{gamma_wr}
\end{figure}

Fig. \ref{gamma_wr} shows the gamma-ray emission from the Perseus
cluster core as produced in the WR model. This gamma-ray emission
is the sum of three different components: neutral pion decay,
ICS-on-CMB ICS emission and bremsstrahlung emission both from
secondary electrons. The dominant component at $E>100$ MeV is the
one due to $\pi^0 \to \gamma \gamma$ decay, and the total flux is
$F(>100 \mbox{ MeV})=2.2\times10^{-8}$ cm$^{-2}$ s$^{-1}$, still
lower than the EGRET upper limit on Perseus, $F(>100 \mbox{
MeV})<3.72\times10^{-8}$ cm$^{-2}$ s$^{-1}$ (Reimer et al. 2003).

\subsection{DM models}

We consider here three neutralino DM models with neutralino masses
$M_{\chi}= 81 (W^+W^-), 40(b {\bar b})$ and $10 (\tau^+ \tau^-)$
GeV, as we already worked out for the analysis of other clusters
(see Colafrancesco \& Marchegiani 2009).

For each neutralino model we consider a radial DM density profile as given by
\begin{equation}
 g_{DM}(r)=\exp [-(2/\alpha)((r/r_c)^\alpha-1)]
 \label{eq.prof.dm}
\end{equation}
(Navarro et al. 2004), with $\alpha=0.17$ and $r_c$ equal to the
core radius of 200 kpc (the larger core radius of the
thermal gas density distribution found by Churazov et al. 2003). We
assume that this DM radial profile extends out to the same region
used in the WR model, $R\sim0.215$ Mpc.
The spectrum of the DM source function for the secondary electrons has,
consequently, a radial distribution $\propto g_{DM}^2(r)$.

To derive the equilibrium spectrum of these secondary electrons in
Perseus we consider the role of the dominant energy loss
mechanisms. These are ICS losses against CMB photons and
synchrotron losses for electrons with energy larger than a few
hundreds MeV (notice that synchrotron losses for magnetic fields
less than 3 $\mu$G, are negligible with respect to the ICS
losses), while at low energies ($\simlt 150$ MeV) the dominant
energy loss mechanisms are Coulombian interactions with the IC gas
particles.\\
For this reason the final spatial distribution of secondary
electrons is proportional to $g_{DM}^2(r)$ at high energies ($>
150$ MeV) and proportional to $g_{DM}^2(r)/n_{th}(r)$ at low
energies ($<150$ MeV).\\
If the magnetic field is larger than $\sim3$ $\mu$G, then the
dominant energy loss mechanism at high energy is synchrotron
emission and, consequently, the radial distribution of the
secondary electron equilibrium spectrum is $\propto
g_{DM}^2(r)/g_B^2(r)$, where $g_B(r)$ is the radial distribution
of the magnetic field intensity.

\begin{figure}[ht]
\begin{center}
 \epsfig{file=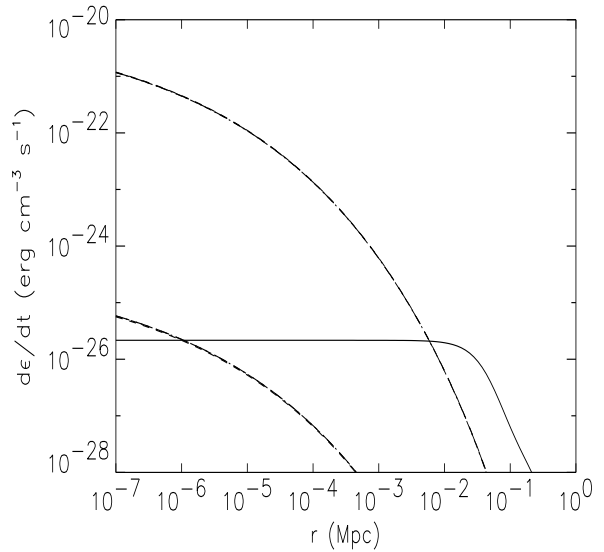,height=8.cm,width=8.cm,angle=0.0}
\end{center}
 \caption{\footnotesize{The heating rate induced by DM-produced secondary
 electrons with a normalization given by the condition that i) the
 heating rate and the cooling rate equal at $r=10^{-6}$ Mpc (lower curves)
 , and ii) that their volume integral within $r=10^{-2}$ is equal (upper curves).
 We show different DM models with $M_\chi=81$ GeV
(long-dashes curves), 40 GeV (short dashes), and 10 GeV
(dot-dashes curves); continuous curve represents the IC gas
cooling due to thermal bremsstrahlung.}}
 \label{heating_dm}
\end{figure}

The value of the DM annihilation cross-section directly determines
the normalization of the pions and of the secondary electron
density and hence of the gamma-ray emission. Since the value of
this cross-section is not know, we fix it by setting the value of
the heating rate due to secondary electrons compared to the value
of the bremsstrahlung cooling of the IC gas in Perseus.
Specifically, we used two different criteria:
\textit{i)} the local heating rate is equal to the local cooling
rate in the center of the cluster, e.g. at $r=10^{-6}$ Mpc;
\textit{ii)} the heating rate, integrated in the volume within
$r=10^{-2}$ Mpc, is equal to the cooling rate integrated in the
same region.
Fig. \ref{heating_dm} shows the comparison between the cooling
rate  and the heating rate due to DM annihilation as normalized
according to the previous methods. The normalization method {\it
ii)} yields a normalization of the secondary particle density and
of the gamma-ray emission larger by a factor $\sim10^4$.
As we already noticed in Sect. 4.1 for the WR model, there can be
various heating mechanisms at work in the cluster atmosphere.
Therefore, the density of relativistic particles produced in DM
annihilation (whose value depends, for a fixed neutralino mass, on
the DM annihilation cross-section) that is set by our analysis can
be considered as an upper limit.
We also stress that, as in the WR model, we assume a spherically
symmetric radial distribution that produces a heating rate that
has a regular spherical symmetry. As discussed in Sect.4.1, the
results of our analysis can be considered as a reasonable
approximation of the overall cluster physical mechanisms.

\begin{figure}[ht]
\begin{center}
 \epsfig{file=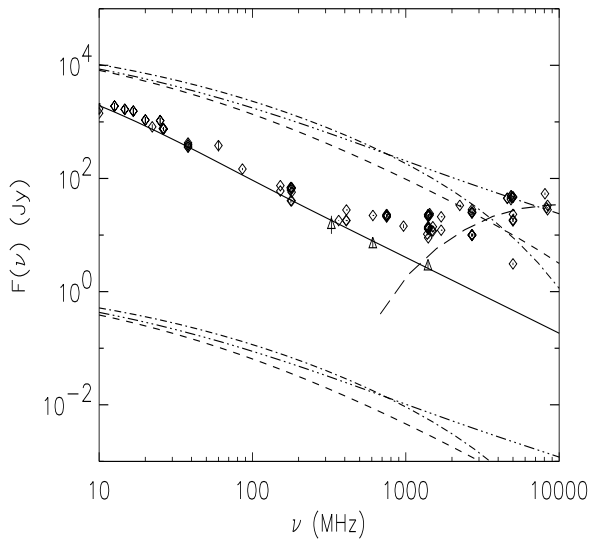,height=8.cm,width=8.cm,angle=0.0}
\end{center}
 \caption{\footnotesize{Radio halo emission from secondary electrons
produced from DM annihilation for a uniform magnetic field of 4
$\mu$G and normalization fixed by the condition that the electron
heating rate is equal to the cooling rate at $r=10^{-6}$ Mpc
(lower curves) or in the volume within $r=10^{-2}$ Mpc (upper
curves). We show predictions for the three DM models with
$M_\chi=81$ GeV (three dots - dashes), 40 GeV (short dashes) and
10 GeV (dot-dashes); the solid thick curve shows the radio
emission expected in the WR model. We also show the SSC model for
NGC 1275 (lomg dashes; component 1 in Tab. \ref{tab.1}). Data are
from Gitti et al. (2002) (triangles) and from NED (diamonds).
 }}
 \label{radio_dm}
\end{figure}

Fig. \ref{radio_dm} shows the comparison between the diffuse radio
emission due to DM-produced secondary electrons and due to the
secondary electrons in WR models (see Sect.2 above) assuming the
same IC magnetic field ($B=4$ $\mu$G costant). The radio halo flux
obtained in WR models is intermediate between the fluxes produced
in the DM models with the two different normalization condition
used here.

\begin{figure}[ht]
\begin{center}
 \epsfig{file=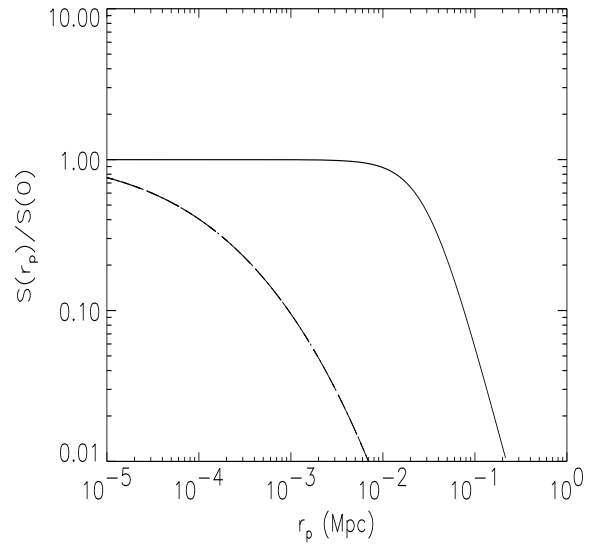,height=8.cm,width=8.cm,angle=0.0}
\end{center}
 \caption{\footnotesize{Brightness profile of the radio halo
 produced by secondary electrons in DM models under normalization
 of the heating rate at the cluster center. We show the
 predictions of DM models with $M_\chi=81$ GeV (log dashes), 40 GeV (short dashes)
 and 10 GeV (dot-dashes); solid curve is the best-fit to the radio halo brightness
 at 1.4 GHz found by Pfrommer \& En\ss lin (2004).
 }}
 \label{brilrad_dm}
\end{figure}

Fig. \ref{brilrad_dm} shows the radio halo brightness profile in
DM models considered here compared to the best-fit of the Perseus
radio halo brightness profile derived by Pfrommer \& En\ss lin
(2004): it is clear that the radio halo brightness profile
produced only by the smooth DM component is much more concentrated
than the observed brightness profile. Such disagreement could be
solved by assuming either a profile of the magnetic field that
increases towards the outer regions of the cluster (a quite
unlikely solution, as discussed by Colafrancesco et al. 2005) or,
more likely, by the presence of DM substructures that can provide
a larger radio brightness in the outer parts of the cluster (as
discussed extensively by Colafrancesco, Profumo \& Ullio 2006).

\begin{figure}[ht]
\begin{center}
 \epsfig{file=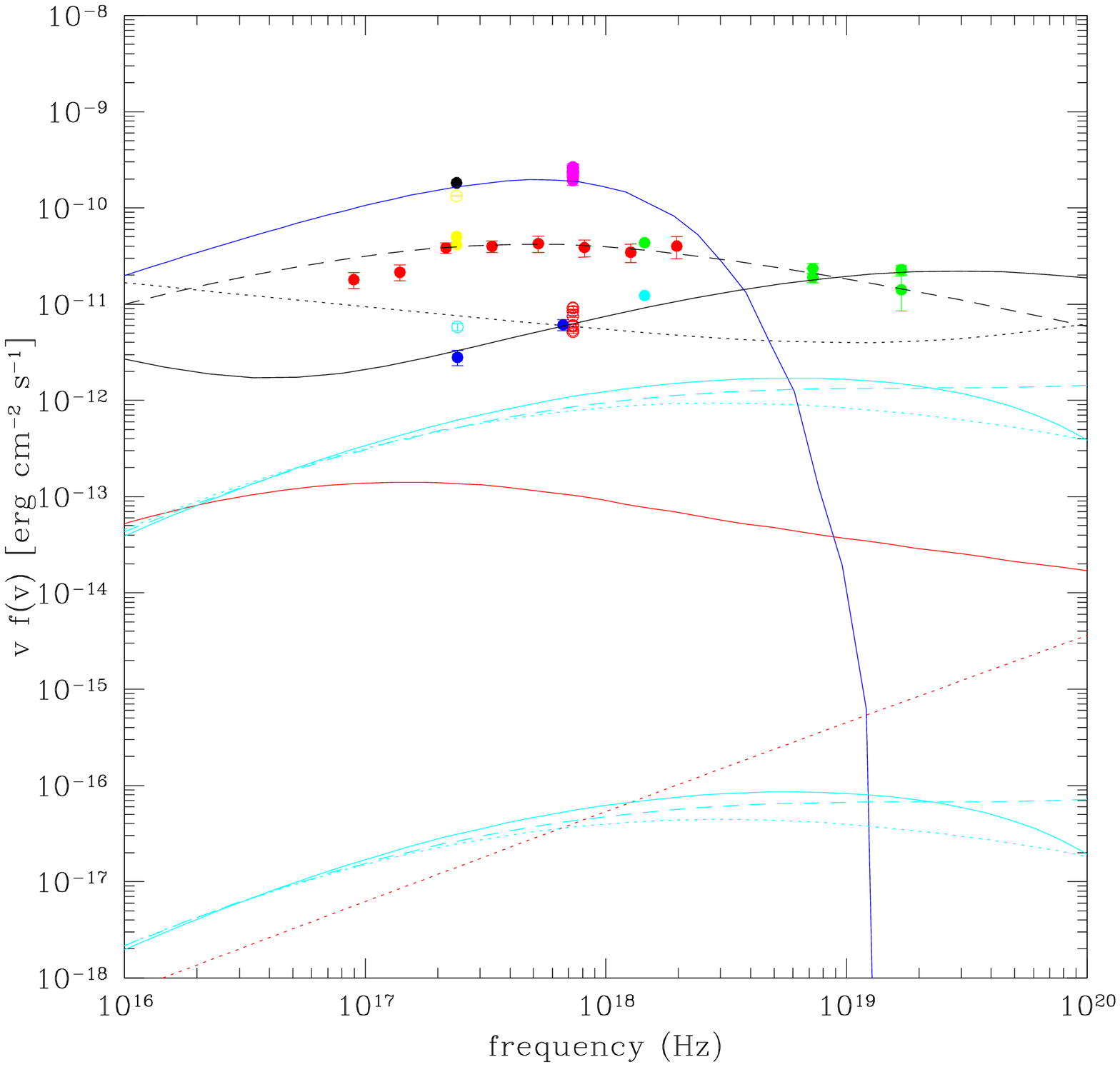,height=8.cm,width=8.cm,angle=0.0}
\end{center}
 \caption{\footnotesize{HXR ICS-on-CMB emission from secondary electrons
produced by DM with normalization set by the condition to have the
local heating rate equal to the cooling rate at $r=10^{-6}$ Mpc
(lower cyan curves) or the integrated ones equal within
$r=10^{-2}$ Mpc (upper cyan curves). Predictions are for DM models
with $M_\chi=81$ GeV (dashes), 40 GeV (dotted) and 10 GeV (solid);
the red curves show the HXR ICS (solid) and bremsstrahlung
(dotted) emission predicted in the WR model, while the blue solid
curve shows the thermal bremsstrahlung emission from the cluster.
The components 1 (black solid), 2 (black dotted) and 3 (black
dashed) of the SSC model for NGC 1275 are also shown for comparison
(see Tab. \ref{tab.1} for details). Data are from NED (see caption
of Fig.\ref{sed_1275_3c_xray} for details).
 }}
 \label{hxr_dm}
\end{figure}

In Fig. \ref{hxr_dm} we compare the X-ray spectrum of the thermal
IC gas with the spectra produced by ICS-on-CMB in the WR model and
in DM models. The flux in the 2--10 keV produced by ICS-on-CMB in
the three DM models, and for the highest normalization level here
considered, are $1.6\times10^{-12}$, $1.4\times10^{-12}$ and
$2.0\times10^{-12}$ erg cm$^{-2}$ s$^{-1}$, for $M_\chi=81$, 40
and 10 GeV, respectively. In all cases, these fluxes are lower
than the flux derived by  Sanders et al. (2005) that is
$6.3\times10^{-11}$ erg cm$^{-2}$ s$^{-1}$.

\begin{figure}[ht]
\begin{center}
  \epsfig{file=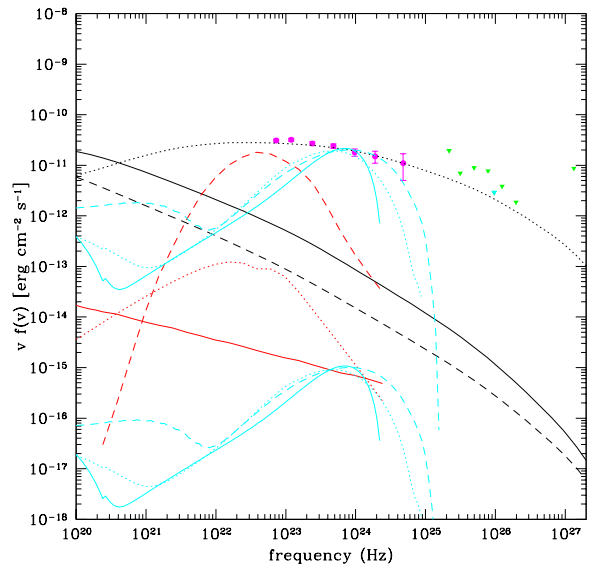,height=8.cm,width=8.cm,angle=0.0}
\end{center}
 \caption{\footnotesize{
The Multiple-component SED of NGC1275 in the gamma-ray frequency
range. The three SED components have parameters listed in
Tab.\ref{tab.1}: components 1 (black solid), 2 (black dotted) and
3 (black dashed). The cluster gamma-ray emission components are
also shown (see caption of Fig.\ref{hxr_dm} for details). Data are
from {\it Fermi} (red points) and the upper limits are from MAGIC
(green) and WHIPPLE (cyan).}}
 \label{sed_1275_perseus_gamma}
\end{figure}

The different components of gamma-ray spectra (i.e. the $\pi^0 \to
\gamma \gamma$, ICS-on-CMB emission and bremsstrahlung from
secondary electrons) expected from the DM models and from the WR
model are shown in Fig. \ref{sed_1275_perseus_gamma}. The
gamma-ray emission from both DM models with the highest
normalization and from the WR model are  below (a factor 3 to 10
from 300 MeV to 8 GeV) the gamma-ray emission detected by {\it
Fermi} towards NGC 1275 and below the EGRET upper limit (see Figs.
11 and 13). These results are consistent with the findings of the
first preliminary associations of galaxy clusters with
unidentified EGRET sources (Colafrancesco 2002). We note that the
spectra shown are within reach of the {\it Fermi} sensitivity with
integration over long exposure times of $\sim 5$ years. The
gamma-ray spectra predicted in DM models with the lowest
normalization here considered are instead much below (a factor
$\sim 10^3$) the 1 yr {\it Fermi} sensitivity and therefore hardly
detectable in the next future.

\subsection{Pure leptonic model}

We want to explore now the consequences of the presence of a
diffuse population of non-thermal relativistic electrons in the
atmosphere of the Perseus cluster without making assumptions on
their origin.
We assume specifically a relativistic electron distribution with a
power-law spectrum $N_e\propto E^{-p}$, with $p=3.7$ such that the
spectral shape of the mini radio halo in Perseus can be
reproduced, and we normalize the density of these electrons in
order to fit the EGRET gamma-ray upper limit. This assumption is
equivalent to find the maximum density of non-thermal electrons
allowed by the gamma-ray upper limit measured by EGRET, and
therefore such case allows us to determine the upper limit on the
Hard X-Ray emission and in turn, combining the radio and HXR data,
the minimum value of the magnetic field of the Perseus atmosphere.
As for the normalization of the electron spectrum, we considered
two different cases:\\
i) the electron density is normalized in such a way to reproduce
the EGRET gamma-ray upper limit with their non-thermal
bremsstrahlung emission. This condition implies that the electron
spectrum must have a break at an energy $E\sim 80$ MeV in order
to have a heating rate smaller than the thermal bremsstrahlung
cooling at the cluster center. With such a spectrum, the gamma-ray
emission produced by these electrons is dominated up to $E\sim4$
GeV by the bremsstrahlung emission component.
The multi-frequency behaviour of this electron spectrum implies
that in the $2-10$ keV band the relativistic electrons produce a
ICS-on-CMB flux of $6.6\times10^{-12}$ erg cm$^{-2}$ s$^{-1}$,
smaller than the flux $6.3\times10^{-11}$ erg cm$^{-2}$ s$^{-1}$
estimated by Sanders et al. (2005) from the center of Perseus. The
same electron population might reproduce the mini radio halo flux
of Perseus assuming a central magnetic field $B_0\sim3$ $\mu$G
under the assumption that the radial profile of the magnetic field
is similar to that of the intracluster gas;\\
ii) the electron density is normalized in such a way to reproduce
the EGRET gamma-ray upper limit with their ICS-on-CMB emission:
in this case, the required electron density is $\sim 60$ times
larger than in the previous case and, consequently, the energy
scale of the spectral break required to avoid an excessive heating
of the thermal gas is $E\sim370$ MeV. With such a spectrum, the
gamma-ray flux is still dominated by the bremsstrahlung component
up to $E\sim4$ GeV, and therefore the model in untenable because
the total gamma-ray emission (bremsstrahlung plus ICS-on-CMB)
would be larger than that detected by {\it Fermi}. The same
conclusion is reinforced by the predictions of the multi-frequency
behaviour of such electron spectrum: the ICS flux in the $2-10$
keV band is $4.2\times10^{-10}$ erg cm$^{-2}$ s$^{-1}$, larger
than that measured, and the central value of the magnetic field
required to fit the mini radio halo spectrum is $\sim0.5$ $\mu$G,
much lower than the values usually derived for cool core clusters.

In conclusion, the density of the relativistic electrons must be
normalized using their non-thermal bremsstrahlung emission, since
this emission dominates on the ICS-on-CMB one. This normalization
implies a lower value of the central magnetic field of $B_0\sim3$
$\mu$G and a lower energy break at $E\sim80$ MeV in the electron
spectrum.

\section{The multi-$\nu$ SED of NGC 1275 and Perseus: a comparison}

\begin{figure}[ht]
\begin{center}
 \epsfig{file=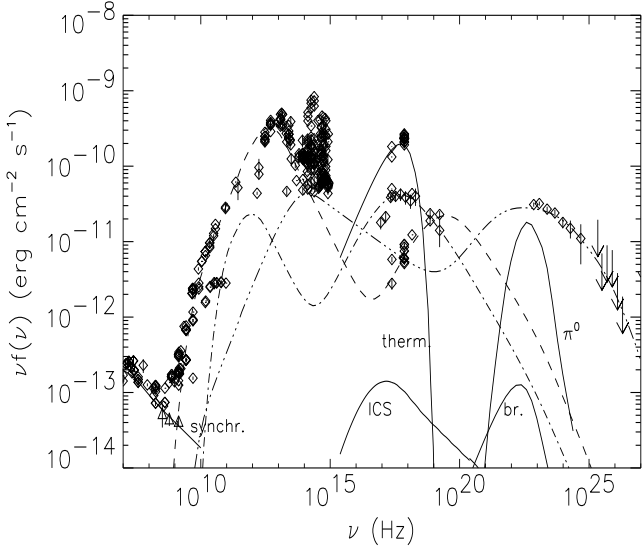,height=8.cm,width=9.cm,angle=0.0}
\end{center}
 \caption{\footnotesize{The three SSC components for NGC 1275 as in Fig.\ref{sed_1275}
 are compared with the various diffuse emission components (as labelled) of the
Perseus cluster in the WR model (solid curves). Data from Gitti
et al. (2002) (triangles) are added w.r.t. Fig.\ref{sed_1275}.
 }}
 \label{sed_1275_cluster}
\end{figure}

Fig. \ref{sed_1275_cluster} shows the three SSC components for NGC
1275 compared to the multi-$\nu$ SED of the diffuse emission of
Perseus in the WR model.
It is interesting to note that the diffuse cluster emission is
stronger than that of the central radio galaxy only at very low
radio frequency ($\simlt 1$ GHz), consistently with the turn over
of the radio data and with the change in their spectral slope, and
differently from the behaviour of the SSC model for the nucleus of
NGC 1275 and of AGNs in general (see Colafrancesco \& Giommi
2006). It is important to notice that at such low frequency radio
measurements have a lower angular resolution that does not allow
to distinguish between the galaxy emission and the diffuse cluster
emission.

Fig.\ref{sed_1275_perseus_gamma} shows the comparison between the
gamma-ray part of the SED of NGC1275 and the gamma-ray emission of
the Perseus cluster.

The cluster SED, both in the WR model and in DM models, are
significantly lower and with significantly different spectral
shape with respect to the SED of NGC 1275 and to the {\it Fermi}
data.\\
The WR model predicts a flux of $2.2\times10^{-8}$ cm$^{-2}$
s$^{-1}$, almost one order of magnitude lower than the NGC 1275
flux observed with {\it Fermi}. In order to reproduce the {\it
Fermi} data, a density of relativistic protons 10 times higher
would be necessary in the WR model, but such high proton density
would provide an excessive heating of the intra-cluster gas.\\
The DM models with their highest mormalization could reach the
{\it Fermi} flux value but could not reproduce the {\it Fermi}
spectrum. We also stress that this normalization produces an
excessive heating rate at the cluster center (see
Fig.\ref{heating_dm}).

The {\it Fermi} data are very well fitted in a CM model with three
components for NGC 1275, in which the blob n.2 (see parameters in
Table \ref{tab.1}) produces all of the emission observed at
gamma-ray energies, while it produces a sub-dominant emission in
the other frequency ranges, apart from the extreme UV and soft
gamma-rays (see Fig.\ref{sed_1275_cluster}). This SED is also
consistent with the existing TeV limits. The gamma-ray historical
variability (see the {\it Fermi} and EGRET data) is also
consistent with a multiple blob scenario in the CM model and
excludes also that the bulk of gamma-ray emission is due to the
diffuse emission from the cluster atmosphere.

In the X-ray frequency range, (see Fig.\ref{hxr_dm}) the various
non-thermal emissions from the cluster are always lower than the
observations obtained with different instruments. The thermal
bremsstrahlung from the cluster is instead clearly observed and is
consistent with the ROSAT and BeppoSAX WFC  observations.
The CM model with three components for NGC 1275 is instead able to
reproduce all other X-ray observations with higher angular
resolution and the historical variability of the source clearly
seen at these frequencies.

Our results indicate that the frequency ranges in which the
diffuse cluster SED is detected, and exceeds the SED of NGC1275,
are those of the low-frequency radio range $\nu\simlt 1$ GHz and
that of the soft X-rays with $E\sim$ keV. The SED observed in all
other frequency ranges is dominated by the central galaxy NGC
1275. The existing data are best explained by a CM model with
three main components in which the first one produces most of the
emission observed from radio to optical frequencies and the lowest
state of the X-ray emission, the second one produces most of the
gamma-ray emission and the third one produces the higher state of
the X-ray observations.

\section{Discussion and conclusions}

In this paper we studied the gamma-ray emission features of the
combined system of the NGC 1275 galaxy living in the core of the
Perseus cluster based on a multi-component model and in
multi-frequency approach. This study has been done with the aim to
assess a strategy that is able to disentangle the gamma-ray
emission of the cluster from that of the galaxy NGC1275.\\
Our analysis provides various constraints on the NGC 1275 emission
properties, on the cluster atmosphere, and on the interaction of
the two sources.


We worked out a CM model for NGC 1275 that is able to reproduce
the {\it Fermi} observations, the yearly (or monthly) variability
of this source in gamma-rays and X-rays, and the different
observed intensity states of its multi-frequency SED (see Figs.
\ref{sed_1275} e \ref{sed_1275_3c_xray}).
We find that the {\it Fermi} detection of NGC 1275 is entirely due
to the emission of a compact ($r \sim 8 \times 10^{-3}$ pc) and
energetic blob filled with an electron population with double
power law spectrum with $\gamma_b \approx 5\times 10^{3}$, with
$\delta \approx 8$ and relatitely high normalization factor $N
\approx 6.8$ cm$^{-3}$ (see Table 1). The SED of this blob
perfectly fits the {\it Fermi} spectrum and is also consistent
with MAGIC and WHIPPLE upper limit at TeV energies.
We also note that the upper limit obtained by VERITAS (Acciari et
al. 2009) is lower than the flux of the high-energy component of
the CM model (see Fig.\ref{1275_tev}). However, Acciari et al.
(2009) pointed that the {\it Fermi} flux measurements simultaneous
with the VERITAS ones are lower by a factor of $\sim1.37$ w.r.t.
the data of Abdo et al. (2009). Therefore, the VERITAS upper
limits are consistent with the fact that they have been measured
while the central source was decreasing its high-energy flux, and
are hence completely consistent with a variable high-E component
of the NGC 1275 SED as predicted by the CM model.


For the surrounding Perseus cluster atmosphere, DM annihilation
models with high normalization are excluded because they produce a
stationary gamma-ray flux of the same order of that observed at $E
\sim 3$ GeV by {\it Fermi} from NGC 1275, and - in addition - they
produce a spectrum of gamma-ray emission that is strongly peaked
at a few GeV, contrary to the spectral shape of the gamma-ray
source detected by {\it Fermi} (see Fig.
\ref{sed_1275_perseus_gamma}).\\
The WR model produces instead a gamma-ray flux always much lower
(by a factor of 5 to more than 10) than the gamma-ray flux from
NGC 1275 observed by {\it Fermi}.\\
For a pure leptonic model, the EGRET upper limit sets an upper
limit to the relativistic electrons density or, combining the
gamma measurements with the radio ones, a lower limit to the
magnetic field. We have verified that the non-thermal
bremsstrahlung emission dominates on the ICS-on-CMB one. For this
model, a lower value of the central magnetic field of $B_0\sim3$
$\mu$G is found, and a lower spectral cutoff of $E \sim 80$ MeV in
the electron spectrum is required by the heating constraints.

\begin{figure}[ht]
\begin{center}
 \epsfig{file=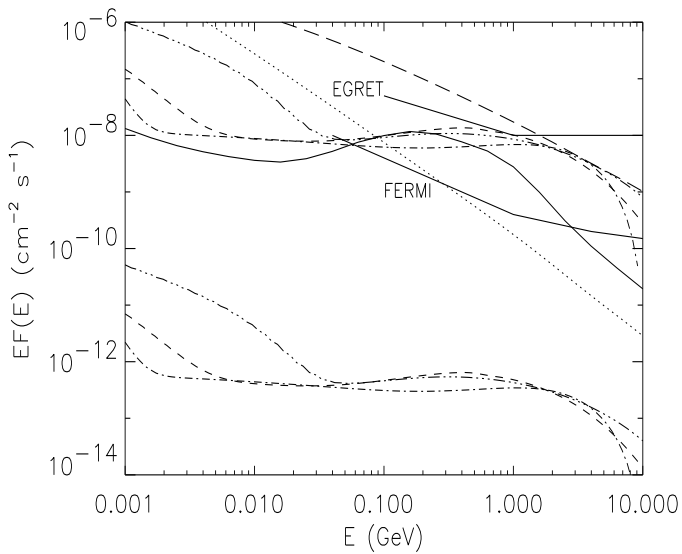,height=8.cm,width=9.cm,angle=0.0}
\end{center}
 \caption{\footnotesize{Gamma-ray emission of the NGC 1275 CM model component
 1 (dotted line) and 2 (long-dashed line), of cluster WR model (thick
 line) and DM models with $M_\chi=$ 81, 40 and 10 GeV (three dots-dashed,
 dashed and dot-dashed lines respectively) with high and low normalization.
 Sensitivity limits at 5$\sigma$ level for 1 yr of observation for EGRET and
{\it Fermi} are also shown.
 }}
 \label{gamma_5sigma}
\end{figure}

\begin{figure}[ht]
\begin{center}
 \epsfig{file=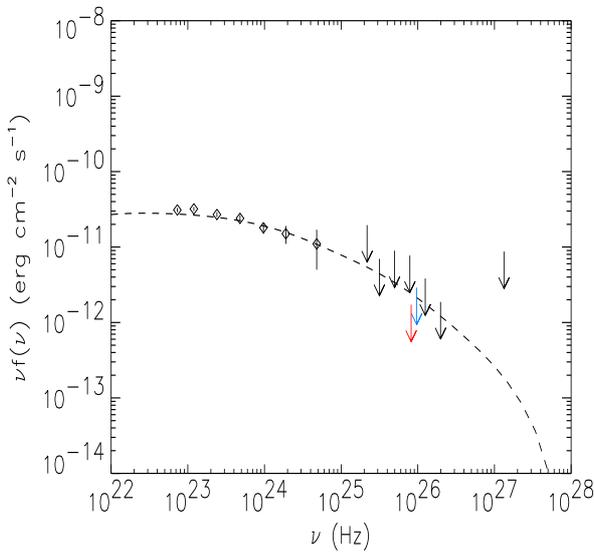,height=8.cm,width=8.cm,angle=0.0}
\end{center}
 \caption{\footnotesize{
A close-up of the high-energy spectral region of the SED of NGC
1275: the dashed line is the component 2 of the CM. Data are from
{\it Fermi} (open circles). The upper limits are from MAGIC
(black), WHIPPLE (blue) and VERITAS (red).}}
 \label{1275_tev}
\end{figure}

Due to the fact that the gamma-ray emission of NGC 1275 is
variable, it is possible to reveal the diffuse gamma-ray emission
from Perseus during a state of low activity of NGC 1275.
In the core of the Perseus cluster, the diffuse gamma-ray flux at
$E> 100$ MeV from the cluster atmosphere (both in the WR model and
in the DM annihilation model with the highest normalization, that
produces an over-heating at the cluster center) could be detected
by {\it Fermi} if NGC 1275 is found in a low-state as that
produced by the first blob in our CM model (component 1 in Table
1, see also Fig.\ref{gamma_5sigma}).
The diffuse gamma-ray flux coming from these diffuse components
might be also marginally detectable by the future CTA experiment
at $E \simgt 10$ GeV.
The gamma-ray flux produced by DM annihilation with the lowest
normalization would be undetectable by both {\it Fermi} and CTA
being its peak flux at $\sim 1$ GeV more than a factor 300 below
the {\it Fermi} (1 yr., $5 \sigma$) sensitivity.

An alternative approach to disentangle the cluster gamma-ray
emission from that of NGC 1275 is to determine the diffuse flux
outside the cluster core where the emission from the central
galaxy is no longer contaminating the measurements. At $E > 1$ GeV
the {\it Fermi} 68\% collecting angle is $\sim 0.6$ deg. that
corresponds to $\sim 786$ kpc from the cluster center: thus {\it
Fermi} can have the possibility to resolve and detect the diffuse
gamma-ray flux coming from the outer corona of the Perseus cluster
atmosphere at distances $r \simgt 800$ kpc from the center of the
cluster. The gamma-ray flux from Perseus evaluated in the corona
between 0.786 and 1 Mpc (i.e., approximately the maximum radius to
which X-ray thermal emission is observed; see, e.g., Furusho et
al. 2001) is $F(> 1 \mbox{ GeV}) \approx 1.2 \times 10^{-10}$
cm$^{-2}$ s$^{-1}$ in WR model, which can be marginally detectable
from {\it Fermi}. Therefore, such a strategy could provide a
realistic case to detect the diffuse gamma-ray emission from the
Perseus cluster.\\
This argument does not apply to the case of DM models because in
such a case the gamma-ray flux comes mainly from the inner cluster
region due to the $\propto n_{DM}^2(r)$ dependence of the
annihilation flux. This implies that even a spatially resolved
observation of the Perseus cluster in gamma-rays cannot provide
information on the DM models since the relative gamma-ray
brightness is not observable at large angular distances from the
central galaxy.

As for the full multi-frequency approach, we find that the diffuse
emission produced in the cluster core is much less than that of
the central galaxy (Fig.\ref{sed_1275_cluster}), with the
exception of two frequency windows: the radio frequency region at
$\nu \simlt 1$ GHz (Fig.\ref{radio_wr}) in which the diffuse mini
radio halo of Perseus dominates the emission, and the soft X-ray
band ($E\sim$ keV), at which the thermal emission from the cluster
gas peaks (Fig.\ref{hxr_wr}). The dominance of the cluster diffuse
emission in these frequency ranges is a well known result.\\
The WR model is the one that better fits both the mini radio halo
features (requiring a magnetic field radial profile almost
constant in the cluster core $R\leq215$ kpc see
Fig.\ref{radiobril_wr} and with a value of $B\sim4$ $\mu$G,
consistently with Faraday Rotation measurements, see Carilli \&
Taylor 2002) and the flux and temperature distribution of the
diffuse, X-ray emitting plasma.
DM annihilation models produce, instead, a much steeper density
profile towards the cluster center (Fig.\ref{brilrad_dm}), and
might then reproduce the radio halo brightness profile only for a
B-field radial profile which rapidly increases towards the outer
cluster regions.

In summary, the gamma-ray emission from the Perseus cluster is
dominated in its central region by the central radio galaxy NGC
1275, a situation similar to those found for the first preliminary
evidence found with EGRET of gamma-ray emitting clusters hosting
powerful radio galaxies (see the previous conclusions by
Colafrancesco 2002). This seems to be the most promising case to
observe galaxy cluster in gamma-rays at the moment, even though
the expected detectable number of clusters is quite limited to a
few specific cases.
These conclusions are certainly quite different from those of
other (accreting and merging cluster) models in which it has been
predicted that the operating gamma-ray missions should have
already discovered a significant number (i.e., 5 to 7 clusters
with AGILE, and about 20 clusters with {\it Fermi}) after 1 yr.
operation (see, e.g., Blasi, Gabici \& Brunetti 2007).\\
The diffuse gamma-ray flux from the Perseus cluster can be
detected in two regions: i) in the cluster core, for both DM and
WR models, when the gamma-ray emission from NGC1275 is found in a
low activity state; ii) in the outer cluster region at $r \simgt
800$ kpc only for the WR model.

In conclusion, our results show that a simultaneous study of the
various emission mechanisms that produce diffuse gamma-rays from
galaxy clusters and the study of the emission mechanisms that
produce gamma-rays from active galaxies residing in the cluster
atmospheres is absolutely crucial first to disentangle the
spectral and spatial characteristics of the gamma-ray emission and
secondly to assess the optimal observational strategy in the
attempt to reveal the still elusive diffuse gamma-ray emission
widely predicted for the atmospheres of large-scale structures.

\begin{acknowledgements}
We acknowledge the extensive use of the multi-frequency data
obtained from the ASI-ASDC and available on line from its web site
http://www.asdc.asi.it .  Historical data for NGC1275 have also
been taken by the NASA/IPAC Extragalactic Database.

\end{acknowledgements}


\end{document}